\documentclass[a4paper]{jpconf}
\usepackage{graphicx}
\begin{document}
\title{Magnetism of $3d$ transition metal atoms on W(001):
submonolayer films}

\author{M Ondr\'{a}\v{c}ek$^1$, J Kudrnovsk\'{y}$^1$, I Turek$^2$ and F M\'{a}ca$^1$}

\address{$^1$ Institute of Physics, Academy of Sciences of the Czech Republic, Prague, CZ}
\address{$^2$ Institute of Physics of Materials, Academy of Sciences of the Czech Republic, Brno, CZ}
\ead{ondracek@fzu.cz%, kudrnov@fzu.cz, maca@fzu.cz, turek@ipm.cz
}

\begin{abstract}
We have investigated random submonolayer films of $3d$ transition metals on W(001).
The tight-binding linear muffin-tin orbital method combined with the coherent potential approximation
was employed to calculate the electronic structure of the films.
We have estimated local magnetic moments and the stability of different magnetic structures,
namely the ferromagnetic order, the disordered local moments and the non-magnetic state,
by comparing the total energies of the corresponding systems.
It has been found that the magnetic moments of V and Cr decrease and eventually disappear with decreasing coverage.
On the other hand, Fe retains approximately the same magnetic moment throughout the whole concentration range
from a single impurity to the monolayer coverage.
Mn is an intermediate case between Cr and Fe since it is non-magnetic
at very low coverages and (ferro)magnetic otherwise.
\end{abstract}

\section{Introduction}
Magnetic properties of thin films on solid state surfaces strongly depend
on surface chemical composition, crystal structure, surface crystallographic orientation, film thickness, interatomic distances, etc.
The combined effects of all these conditions on the surface magnetism may lead
to unexpected results.
The magnetic ordering of $3d$ transition metal monolayers deposited on the 
bcc(001) tungsten surface is an example.

Experiments indicated that Fe monolayer on W(001) surface
has zero net magnetization (see \cite{wulfhekel} and references therein).
The question whether the layer was really non-magnetic or rather antiferromagnetic was resolved by
Kubetzka {\it et al.} \cite{kubetzka} using the spin-polarized scanning tunneling microscope.
The microscopic image clearly shows the $c(2\times 2)$ antiferromagnetic order of the Fe/W(001) monolayer. 
A subsequent theoretical study \cite{ferriani} confirmed that Fe and Co monolayers tend to be antiferromagnetic
on the W(001) surface, while V, Cr, and Mn monolayers tend to the ferromagnetic order. 
An opposite trend is observed for films on the W(110) surface,
where V, Cr and Mn monolayers are antiferromagnets and Fe and Co monolayers order ferromagnetically.
Similar magnetic ordering as on W(110) takes place on other substrates, e.g.,
on the Cu, Ag, and Pd (001) surfaces (see references in \cite{ferriani}).

The above mentioned studies focused on monolayer films.
This paper extends the investigation to the case of partial coverage ($\theta$) of the (001) tungsten surface.
Disordered submonolayer films may occur, e.g., during the epitaxial growth of the monolayers.
We estimate the local (atomic) moments in the $3d$ metal overlayers and determine their mutual alignment
as a function of coverage.

\section{Method}
We have performed spin-polarized calculations within the local density approximation (LDA)
to determine the electronic structure of the studied systems.
We have applied the tight-binding linear muffin-tin orbital (TB-LMTO) method
in the framework of the atomic sphere approximation (ASA).
The exchange-correlation potential of Vosko, Wilk and Nusair \cite{vwn} was used.
In addition to the standard ASA, a dipole surface barrier was included.
The electronic structure of overlayers deposited on effectively semi-infinite substrate
was treated in the framework of the surface Green function formalism,
which facilitates the inclusion of realistic boundary conditions.
We assumed random distribution of overlayer atoms on the surface,
omitting any possible clustering effects.
The random nature of the adsorbate layer was described by means of the coherent potential approximation (CPA)
(for more details, see \cite{turek}).

The experimental lattice constant of tungsten, $a=$ 3.16~\AA, was used.
%The surface Brillouin zone was sampled by $30\times 30$ $k$-points.
The self-consistent electronic structure calculations were performed on an eight-layer slab:
it consisted of four tungsten layers, one adsorbate layer
and three empty layers that mimicked the vacuum above the surface.
The adsorbate overlayer contained randomly occupied adsorption positions
for the $3d$ metal atoms.
The slab was matched to a semi-infinite vacuum region on one side and to semi-infinite bcc tungsten
with frozen bulk potentials on the other side.

The calculations have been carried out both with and without considering surface relaxations.
In the latter case, the adsorbate atoms are situated in bcc positions
that form an ideal continuation of the tungsten substrate.
To include the relaxation, we have reduced distances between the adsorbate and the first tungsten layer.
We have used the values of the distances found for monolayers \cite{ferriani}.
The first interlayer distance was reduced with respect to the bulk tungsten by
8.4\,\% for V, 14.4\,\% for Cr, 4.7\,\% for Mn, 13.7\,\% for Fe and 22.1\,\% for Co,
irrespective of the magnetic state and coverage.

We have compared the zero-temperature stability of non-magnetic (NM), ferromagnetic (FM), 
and disordered local moment (DLM) states of the films in terms of their total energies.
The DLM state can be considered as a dynamical antiferromagnetic state:
the spin is oriented randomly at each site so that the net effect is zero magnetization.
The DLM state can be naturally described within the CPA (see \cite{turek} for more details).
Lower energy of the DLM state as compared to the FM state indicates a tendency to the antiferromagnetic ordering.
%The ordered $c(2\times 2)$ antiferromagnetic state itself cannot be realized in the disordered incomplete overlayers.
In the NM state, the local magnetic moments of all atoms are zero.

\section{Results and discussion}
Our results for the total energy are consistent 
with the conclusions of \cite{ferriani} for full monolayers:
The FM state is the ground state in the case of V, Cr and Mn monolayers,
while the DLM state has lower energy for Fe monolayers.
The DLM state has lower energy than the FM and NM states for the unrelaxed Co monolayer, too.
No stable DLM or FM state of Co overlayers is found if we account for the relaxation,
but the $c(2\times 2)$ antiferromagnetic state turns out to be the ground state even for the relaxed geometry,
in agreement with \cite{ferriani}.

\begin{figure}[h]
\begin{minipage}{18pc}
\includegraphics[width=18pc]{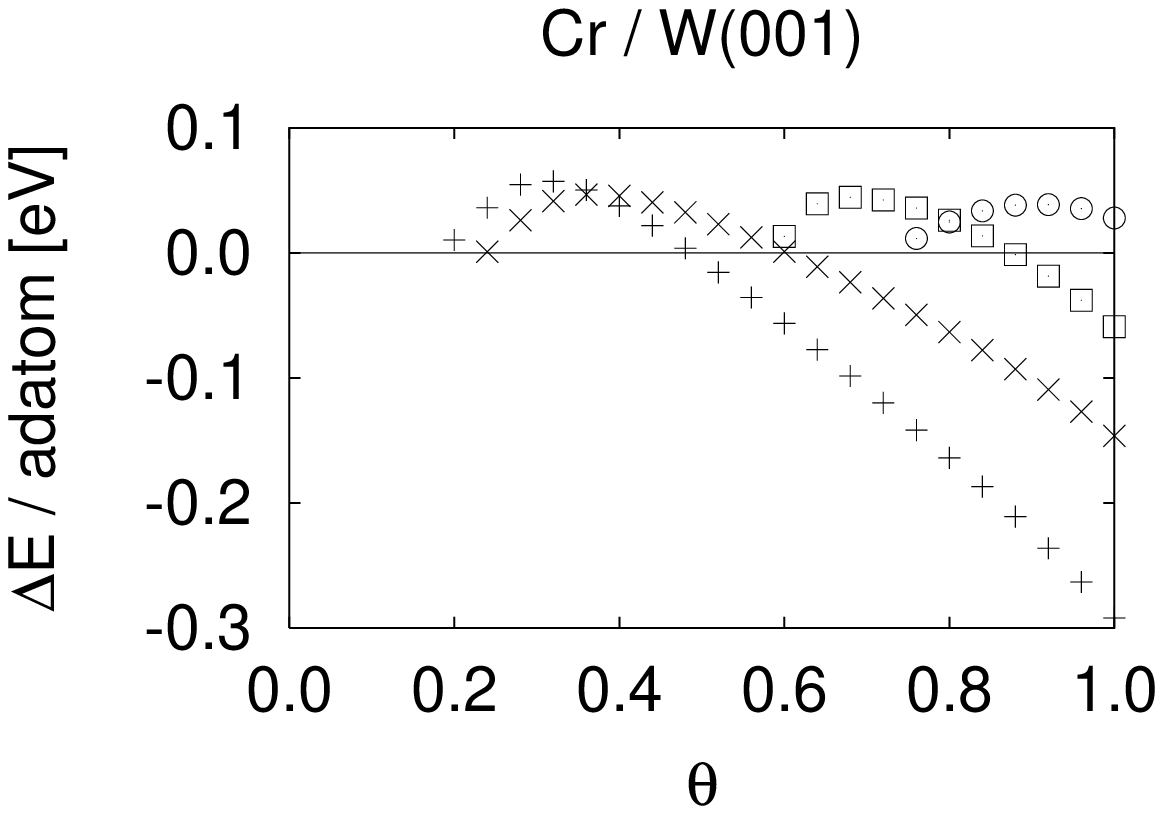}
\end{minipage}
\hspace{2pc}
\begin{minipage}{18pc}
\includegraphics[width=18pc]{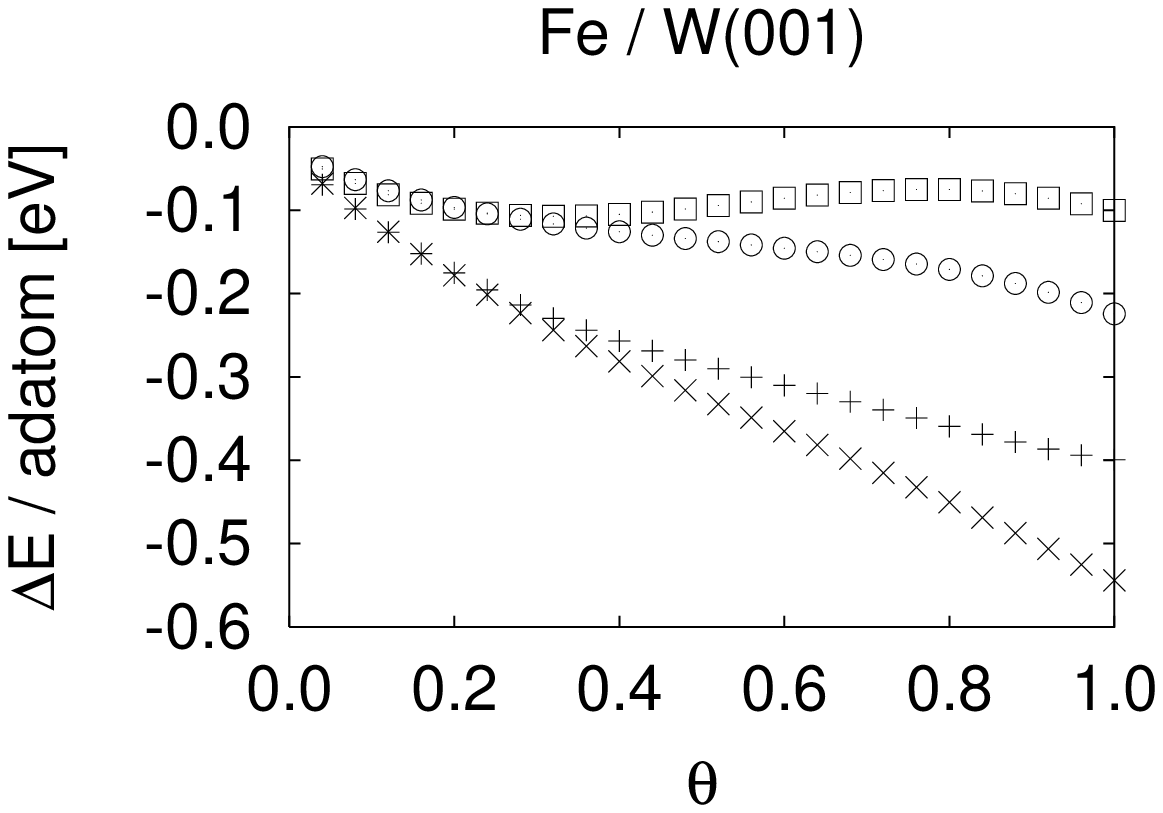}
\end{minipage}
\caption{\label{Ene}
Total energy differences per adsorbate atom
between different magnetic states of incomplete Cr (left) and Fe (right) layers on W(001)
as a function of the coverage: 
FM state, no relaxation ($+$); DLM state, no relaxation ($\times$);
FM state, relaxed (\opensquare)
and DLM  state, relaxed (\opencircle).
The total energy of corresponding NM states is taken as zero.}
\end{figure}

\begin{figure}[h]
\begin{minipage}{18pc}
\includegraphics[width=18pc]{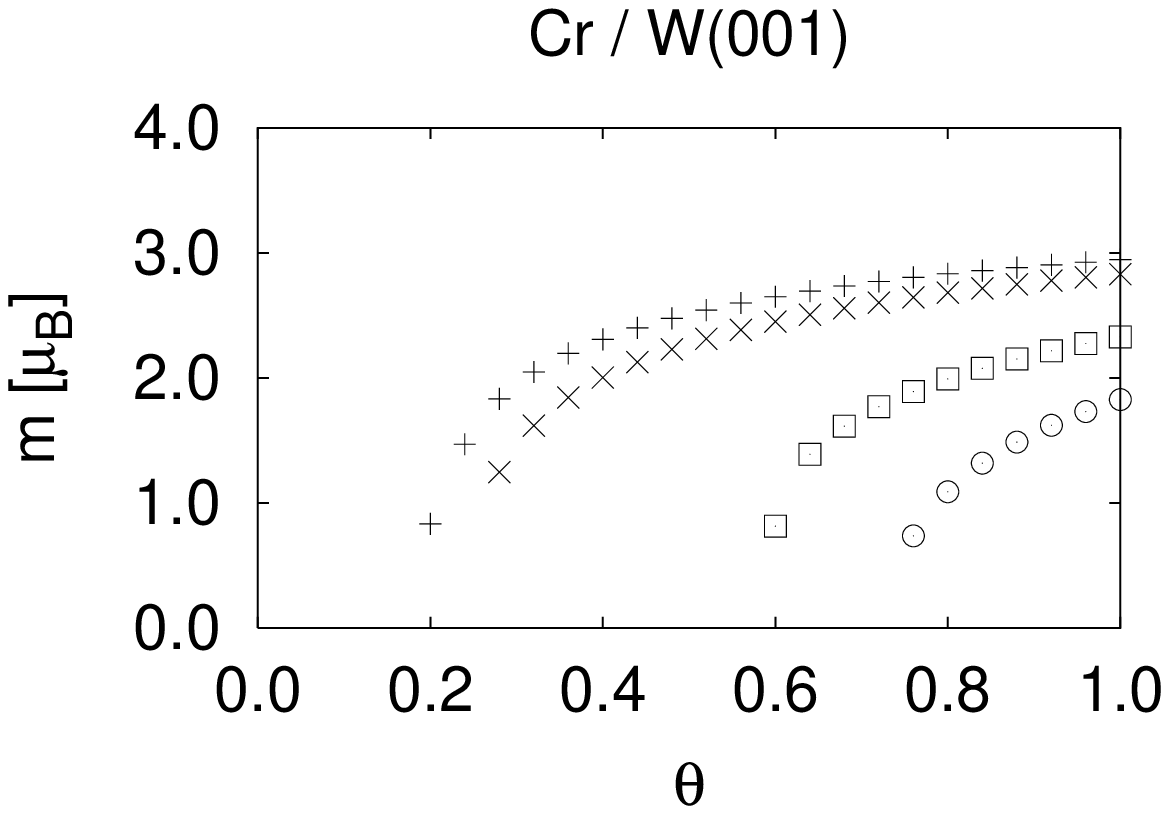}
\end{minipage}
\hspace{2pc}
\begin{minipage}{18pc}
\includegraphics[width=18pc]{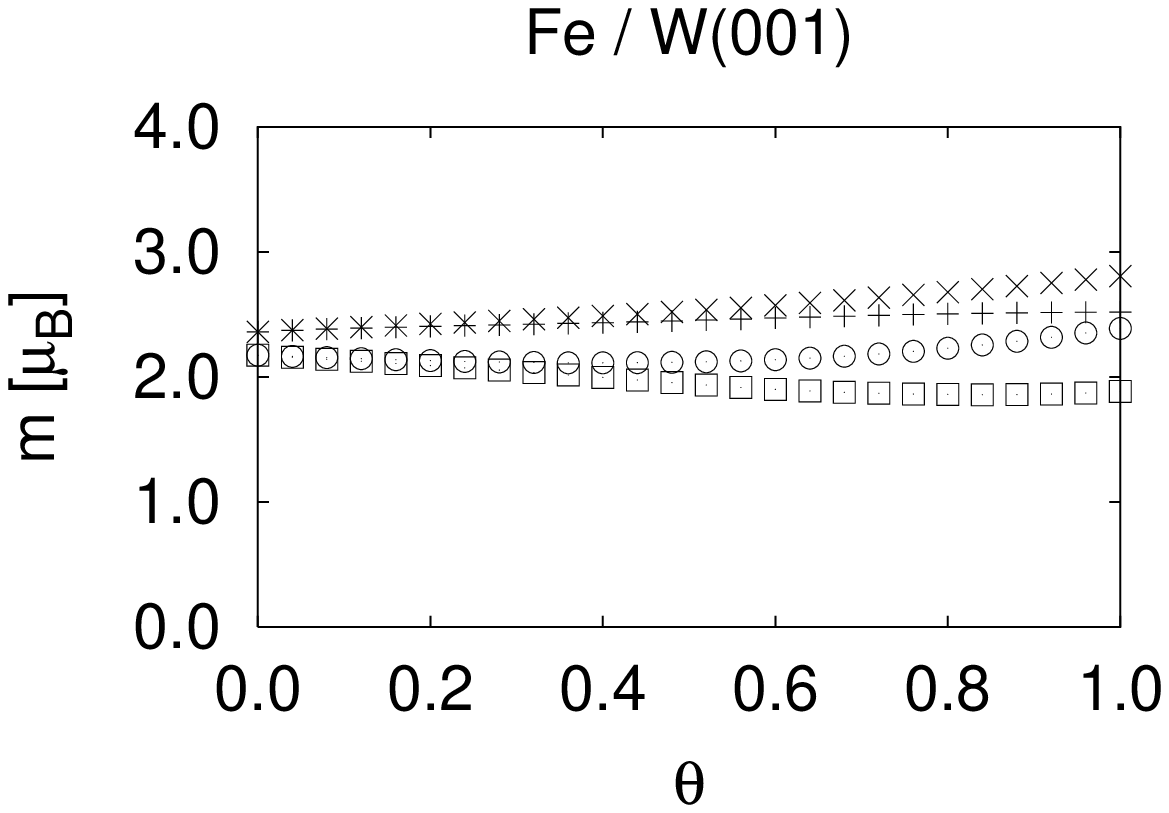}
\end{minipage}
\caption{\label{MagMom}Local magnetic moments of Cr (left) and Fe (right) atoms on the W(001) surface
as a function of the coverage:
unrelaxed FM overlayer ($+$), unrelaxed DLM overlayer ($\times$), 
relaxed FM overlayer (\opensquare) and relaxed DLM overlayer (\opencircle).}
\end{figure}

In submonolayer films, magnetic moments of V, Cr and Mn decrease with decreasing coverage.
The NM state becomes below certain coverage energetically more favorable as compared to the FM state.
For even lower coverage, the atomic moments in the FM and DLM states of V and Cr films 
collapse to zero and no self-consistent magnetic solution can be found
(for Cr, see Figs.~\ref{Ene} and \ref{MagMom}, left panels).
For Mn, magnetic states exist even at the lowest coverages,
although the ground state is non-magnetic there.
The transition between the FM and NM states of V, Cr and Mn overlayers occurs
for unrelaxed geometry at coverages of 0.65, 0.50 and 0.12, respectively.
The critical coverages for this transition in relaxed V, Cr and Mn overlayers are
0.75, 0.88 and 0.25, respectively.

Fe layers are magnetic in the whole range of submonolayer coverages
(see Figs.~\ref{Ene} and \ref{MagMom}, right panels).
The DLM state has lower energy than the FM state at any Fe coverage above $\theta \approx0.20$.
At very low coverages, the dependence of total energy on the magnetic ordering is very weak
because of large adatom distances and the FM and DLM states become nearly energetically degenerate.
The unrelaxed submonolayer Co films exhibit similar behavior as the Fe films,
but we have found no magnetic state for relaxed Co films.
The dramatic change undergone by the Co layers due to surface relaxation results from the fact
their relaxation is the largest among the studied overlayers: over 20\,\% of interlayer distance.
The magnetic states of the Co layers exist only for 
relaxations up to 18\,\% according to our results.
%so the dependence on geometry is very strong for this overlayer.
However, the TB-LMTO method can take into account only small interlayer relaxations reliably. 
From our experience, the upper limit is 15 -- 20\,\%.

The surface relaxation suppresses surface magnetism
due to stronger bonding with substrate that leads to broadened $d$-bands.
Generally, magnetic moments of adatoms are lower for the relaxed geometry as compared to unrelaxed systems
(cf. Fig.~\ref{MagMom}).
The NM--FM transition in V, Cr and Mn films shifts to higher coverages after relaxation,
rendering a wider range of coverages for which these films are non-magnetic 
(cf. Fig.~\ref{Ene}, left panel).

\begin{figure}[h]
\begin{minipage}{18pc}
\includegraphics[width=18pc]{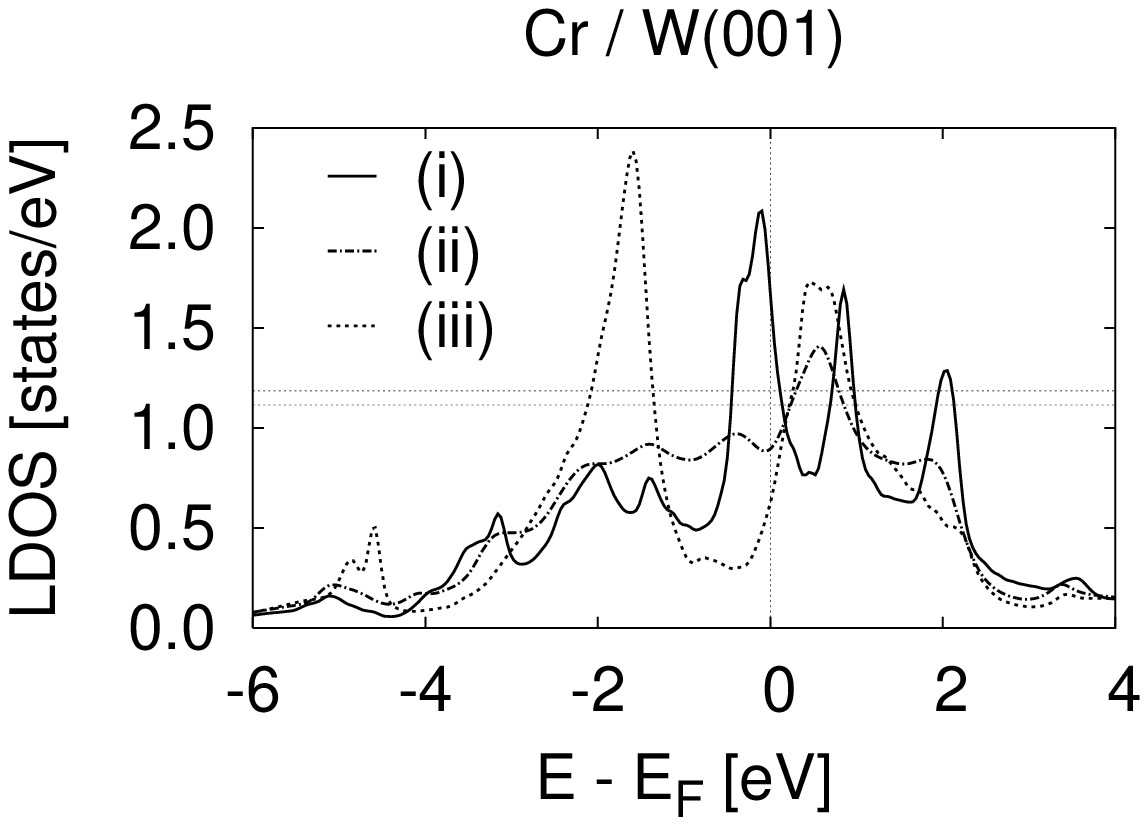}
\end{minipage}
\hspace{2pc}
\begin{minipage}{18pc}
\includegraphics[width=18pc]{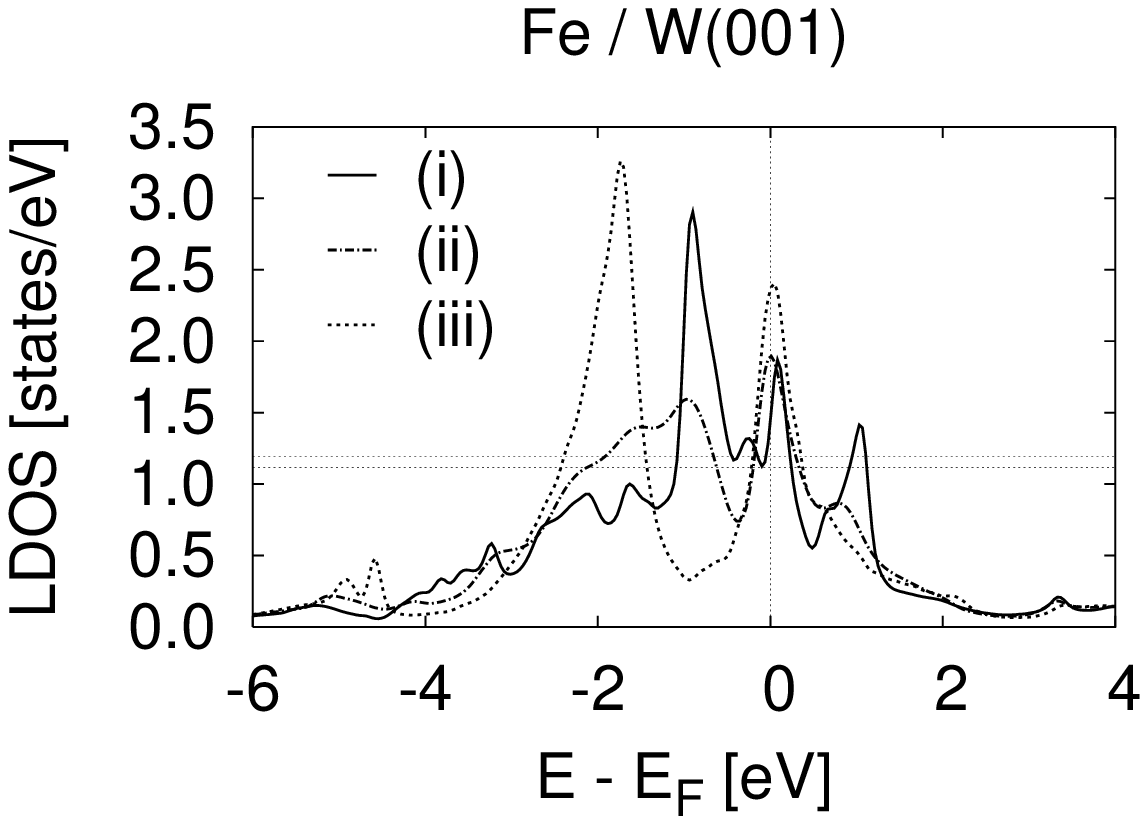}
\end{minipage}
\caption{\label{LDOS}Local density of states per spin at adsorbate atoms (Cr or Fe)
in non-magnetic Cr (left) and non-magnetic Fe (right) relaxed overlayers for different coverages:
(i) full monolayer, (ii) 50\% coverage, (iii) single Cr or Fe atom on the surface.
The horizontal lines illustrate the Stoner criterion (see details in text).}
\end{figure}

The surprising preference for NM states of V, Cr and Mn overlayers at low coverages
(where one would expect enhancement of magnetic moments due to stronger localization of $d$-electrons 
corresponding to the reduced number of neighboring atoms)
can be understood on the basis of NM densities of states.
Local densities of states (LDOS) projected onto the overlayer atoms are shown in Fig.~\ref{LDOS}
for non-magnetic Cr and Fe films.
According to the Stoner criterion \cite{stoner}, a system becomes ferromagnetic if
\begin{equation}
I \cdot n(E_F) > 1,\label{criterion}
\end{equation}
where $n(E_F)$ is the density of states per spin
at the Fermi level of the NM system and $I$ is the so-called Stoner parameter.
The Stoner parameter can be defined using the properties of the corresponding ferromagnetic system
as \begin{equation}
I = \Delta \epsilon_{d} / m ,
\end{equation}
where $m$ is the magnetic moment
(local magnetic moment of the adatom) and $\Delta \epsilon_{d}$ is
the magnetic splitting of electron ($d$-band) energy.
The estimated Stoner parameter, e.g., for Cr and Fe films
is 0.84 -- 0.90~eV.
The parameter varies only slightly with both the adatom type and the coverage
(similar behavior was reported in \cite{feco}).
Two horizontal lines in Fig.~\ref{LDOS} 
indicate the threshold density of states for appearance of the magnetic states
according to Eq.~\ref{criterion}, representing the two above given values of $I$.
There is a peak of LDOS at the Fermi level for both Fe and Cr monolayers,
which ensures that the Stoner criterion is satisfied.
The peak at the Fermi level persists at low coverages for Fe.
For low Cr coverage, however, the peak shifts away from the Fermi level,
leaving the low-coverage Cr adsorbate non-magnetic.
For V adatoms, the changes of the non-magnetic local density of states near the Fermi level with coverage
are similar to the changes observed for Cr layers.
The non-magnetic ground-state of Mn adsorbate at low coverages, on the other hand,
defies the Stoner criterion.
The simple Stoner model is a good approximation for ordered systems,
both three- and two-dimensional (like the monolayers). 
The behavior of magnetic susceptibility for disordered system
is generally more complicated \cite{levin}.
Still, the Stoner criterion turned out to be a good qualitative estimate of magnetism for V, Cr and Fe 
even in the case of incomplete layers.

\section{Conclusions}
We have shown that V, Cr and Mn atoms on the W(001) surface possess no magnetic moment
up to certain coverage,
in contrast to Fe atoms, which are always magnetic on this surface.
It has been found that the magnetic ground state of Fe layers
and the non-magnetic ground states of low-coverage V and Cr layers
can be explained by the Stoner model
in terms of the non-magnetic local density of states.
We have verified that the Stoner parameter depends
only weakly on both the environment and the atom type.
We have confirmed that the atomic moments in magnetic V, Cr and Mn layers favor ferromagnetic order,
%(parallel arrangement)
while the atomic moments of Fe and Co favor antiparallel alignment.
The relaxation of the $3d$ transition metal overlayers on W(001) 
does not change the results qualitatively 
if the change of the overlayer position is
below approximately 15\,\% of tungsten interlayer distance,
which is the case for V, Cr, Mn and Fe films.

\ack
We acknowledge the financial support by COST P19-OC150 
and Grant Agency of the Czech Republic, Grant No. 202/04/583.
We thank Paolo Ferriani for helpful discussion.

\end{document}